\documentclass[pra,tightenlines,showpacs,nofootinbib]{revtex4}

\usepackage{dcolumn,bm,graphicx,amsmath,amsfonts,amssymb}

\begin{document}
\preprint{UNR July 2004--\today }
\title{``Dressing'' lines and vertices in calculations of matrix
elements with the coupled-cluster method and  determination of
Cs atomic properties }
\author{Andrei Derevianko}
\email{andrei@unr.edu}
\affiliation { Department of Physics, University of Nevada, Reno, Nevada 89557}
\author{Sergey G. Porsev}
\affiliation {Department of Physics, University of Nevada, Reno, Nevada 89557}
\affiliation{Petersburg Nuclear Physics Institute, Gatchina,
Leningrad district, 188300, Russia}

\date{\today}

\begin{abstract}
We consider evaluation of matrix elements with the coupled-cluster
method. Such calculations formally involve infinite number of terms and
we devise a method of partial summation (dressing) of the resulting series.
Our formalism is built upon an expansion
of the product $C^\dagger C$ of cluster amplitudes $C$ into a sum of $n$-body
insertions. We consider two types of insertions: particle/hole line
insertion and two-particle/two-hole random-phase-approximation-like insertion. We  demonstrate
how to ``dress'' these insertions and formulate iterative equations.
We illustrate the dressing equations in the case when
the cluster operator is truncated at single and double excitations.
Using univalent systems as an example, we upgrade coupled-cluster diagrams for matrix elements
with the dressed insertions
and highlight a relation to pertinent fourth-order diagrams. We illustrate
our formalism with relativistic calculations of hyperfine constant $A(6s)$ and
$6s_{1/2}-6p_{1/2}$ electric-dipole transition amplitude for Cs atom.
Finally, we augment the truncated coupled-cluster calculations with
otherwise omitted fourth-order diagrams. The resulting
analysis for Cs is complete through the fourth-order of many-body perturbation theory and
reveals an important role of triple and disconnected quadruple excitations.
\end{abstract}


\maketitle

\section{Introduction}
\label{Sec:Intro}
Coupled-cluster (CC) method~\cite{CoeKum60,Ciz66}
is a powerful and ubiquitous technique for solving
quantum many-body problem.
Let us briefly recapitulate general features of the CC method,
so we can motivate our further discussion. At the heart of the CC method lies
the exponential ansatz for the exact many-body wavefunction
\begin{equation}
| \Psi_i \rangle = \exp( T_i ) | 0_i \rangle =
 ( 1+ T_i+ \frac{1}{2!} T_i^2 + \cdots)| 0_i \rangle \, .
\label{Eq:CCparam}
\end{equation}
Here $T_i=\sum_k T_i^{(k)}$ is the cluster operator involving  amplitudes $T_i^{(k)}$
of $k$-fold particle-hole excitations from the reference Slater determinant
$| 0_i \rangle$. The parametrization~\eqref{Eq:CCparam}
is derived from rigorous re-summation of many-body perturbation theory (MBPT) series.
From solving the eigenvalue equation one determines the cluster
amplitudes and the associated energies. While the ansatz
~(\ref{Eq:CCparam}) contains an {\em infinite} number of terms due
to expansion of the exponent, the resulting equations for cluster
amplitudes $T_i^{(k)}$ contain a {\em finite} number of terms. This
simplifying property is unfortunately lost when the resulting
wavefunctions are used in calculations of matrix elements: upon
expansion of exponents the number of terms becomes infinite. Indeed,
consider matrix elements of an operator $Z$, e.g., transition
amplitude between two states
\begin{equation}
\mathcal{M}_{ij}=\frac{Z_{ij}}
{\sqrt{N_i N_j}}
\label{Eq:melGeneral} \, ,
\end{equation}
with normalization $N_i = \langle \Psi_i |\Psi_i \rangle$.
It is clear that both the numerator and denominator have infinite numbers of
terms, e.g.,
\begin{equation}
Z_{ij}= \langle \Psi_i | Z |\Psi_j \rangle =
 \sum_{\lambda=0}^{\infty} \sum_{\mu=0}^{\infty} \frac{1}{\lambda!\mu!}
 \langle 0_i | (T_i^\dagger)^\lambda \,  Z \, (T_j)^\mu | 0_j \rangle \, .
\label{Eq:ZmelSeries}
\end{equation}
In this paper we address a question of partially summing the terms of the above expansion
for matrix elements,
so that the result subsumes an infinite number of terms.

More specifically we are interested in transitions between states of univalent
atoms, such as alkali-metal atoms. There has been a number of relativistic
coupled-cluster calculations for these systems
\cite{BluJohLiu89,BluJohSap91,SafDerJoh98,SafJohDer99,SahGopCha03,GopMerCha02,EliKalIsh94}.
In particular,
calculations~\cite{BluJohLiu89,BluJohSap91,SafDerJoh98,SafJohDer99}
ignore the non-linear terms ($\lambda>1$ and  $\mu>1$) in
the expansion~(\ref{Eq:ZmelSeries});
we will designate this approximation as linearized coupled-cluster (LCC) method.
At the same time, it is well established that for the univalent atoms an
important chain of many-body diagrams for matrix elements
comes from so-called random-phase approximation (RPA).
A direct comparison of the RPA series and the truncated LCC expansion in
Ref.~\cite{DerEmm02} leads to a conclusion that a fraction of the RPA chain is
missed due to the omitted non-linear terms. One of the methods to correct for the missing
RPA diagrams has been investigated in Ref.~\cite{BluJohSap91}. These authors replaced
the bare matrix elements with the dressed matrix elements as prescribed
by the RPA method. Such a direct RPA dressing involved a partial subset of
diagrams already included in the CC method, i.e., it leads to a double-counting of diagrams.
To partially rectify this shortcoming,
the authors of Ref.~\cite{BluJohSap91} have manually removed certain
leading-order diagrams, higher-order terms being doubly counted.
Here we present an alternative infinite-summation scheme for RPA chain that
avoids the double counting and thus a manual removal of the ``extra'' diagrams.

In addition to the RPA-like dressing of the coupled-cluster diagrams for matrix elements,
we consider another subset of diagrams that leads to a dressing of particle
and hole lines in the CC diagrams. The leading order corrections due to
the dressing scheme presented here arise in the fourth order of MBPT,
and in this paper we present a detailed comparison with the
relevant fourth-order diagrams. Finally, we illustrate our approach with relativistic computation of
hyperfine-structure constants and dipole matrix elements for Cs atom.
In addition to dressing corrections we incorporate certain classes of
diagrams from the direct fourth-order MBPT calculation (as in Ref.~\cite{DerEmm02,CanDer04}),
so that the result is
complete through the fourth order. To the best of our knowledge, the reported calculations
are the first calculations for Cs complete through the fourth order of MBPT.

The paper is organized as follows. First, we present a more extensive
discussion of the CC formalism in Section~\ref{Sec:CCformalism}.
Further we dress particle and hole lines in Section~\ref{Sec:ParticleHole},
and discuss RPA-like dressing in Section~\ref{Sec:RPA}.
The present paper may be considered as an all-order extension of
forth-order calculation~\cite{DerEmm02,CanDer04}, and in Section~\ref{Sec:IV}
we present an illustrative comparison
with the IV-order diagrams. Finally, the designed summation schemes are illustrated numerically
in Section~\ref{Sec:Numerics} and the conclusions are
drawn in Section~\ref{Sec:Conclusion}. Unless noted otherwise,
atomic units, $\hbar=|e|=m_e\equiv 1$, are used throughout the paper.
We  follow the convention of Ref.~\cite{LinMor86}
for drawing Brueckner-Goldstone diagrams.

\section{Coupled-cluster formalism for univalent systems}
\label{Sec:CCformalism}
In this Section we specialize our discussion
of the coupled-cluster method
to the atomic systems with one valence electron outside closed
shell core. We review various approximations and summarize the CC formalism for
calculation of matrix elements.

We are interested in solving the atomic many-body problem. The total
Hamiltonian $H$ is partitioned as
\begin{equation}
H=H_0 + G \, ,
\end{equation}
where $H_0$ is the suitably chosen lowest-order Hamiltonian and the residual
interaction $G=H-H_0$ is treated as a perturbation. For systems with one valence
electron outside closed-shell core,
a convenient choice for $H_0$ is the frozen-core ($V^{N-1}$) Hartree-Fock Hamiltonian~\cite{Kel69}.
In the following, we  explicitly specify the state $v$ of the valence  electron,
so that the proper reference eigenstate $| 0_i \rangle$ of $H_0$ is
$| 0_v \rangle =a^\dagger_v | 0_c \rangle$, where pseudo-vacuum state
$| 0_c \rangle$ specifies the occupied core.

For open-shell systems a general CC parametrization reads~\cite{LinMor86}
\begin{equation}
| \Psi_i \rangle = \left\{ \exp( T_i ) \right\} | \Phi_i \rangle
\, ,
\label{Eq:CCparamOpenShell}
\end{equation}
where curly brackets denote normal product of operators.
For univalent system the above ansatz may be simplified to
\begin{equation}
|\Psi_{v}\rangle=
\exp\left(  C\right )\, S_{v} a_{v}^{\dagger}|0_{c}\rangle =
\left(
\sum_{\mu=0}^{\infty} \frac{\left(  C\right)^\mu}{\mu!}
\right) \, S_{v} a_{v}^{\dagger}|0_{c}\rangle \, .
\end{equation}
Here $C$ represents cluster operator involving (single, double, triple, etc.) excitations of
core orbitals
\begin{eqnarray}
\lefteqn{C= C^{(1)} + C^{(2)} + \cdots =} \label{Eq:CoreCluster}
\\
&&\sum_{ma}\rho_{ma}~a_{m}^{\dagger}a_{a}+\frac{1}{2!}\sum_{mnab}\rho
_{mnab}~a_{m}^{\dagger}a_{n}^{\dagger}a_{b}a_{a}+\cdots \, ,
\nonumber
\end{eqnarray}
and $S_{v}$ incorporates additional excitations from the valence state $v$%
\begin{eqnarray}
\lefteqn{S_{v}=1+ S_v^{(1)} + S_v^{(2)} + \cdots } \label{Eq:ValCluster}
\\
&=&1+\sum_{m}\rho_{mv}~a_{m}^{\dagger}a_{v}+\sum_{mna}\rho_{mnav}%
~a_{m}^{\dagger}a_{n}^{\dagger}a_{v}a_{a}+\cdots \, .
\nonumber
\end{eqnarray}
In these formulae and throughout the paper we employ the following labeling convention:
indexes $a,b,c,d$ denote single-particle states occupied in the core $|0_{c}\rangle$
and indexes $m,n,r,s, t$ stand for remaining (virtual/excited) orbitals. In this convention
valence states $v$ and $w$ form a subset of the virtual orbitals. Finally, indexes
$i,j,k,l$ stand for any of the above classes of single-electron orbitals.
In Eqs.~(\ref{Eq:CoreCluster},\ref{Eq:ValCluster}) the cluster amplitudes $\rho_{ij}$
stand for single-particle excitations and $\rho_{ijkl}$ for two-particle excitations,
with an apparent generalization to $k$-fold excitation amplitudes.

Dictated by the computational complexity, in most applications the cluster operator is
truncated at single and double excitations
(CCSD approximation): $C \approx C^{(1)} + C^{(2)}$ and $S_{v} \approx 1+ S_v^{(1)} + S_v^{(2)}$.
A further {\em linearized} (LCCSD) approximation consists in neglecting non-linear
terms in the expansion of exponent in Eq.(\ref{Eq:CCparamOpenShell}), i.e.,
\begin{equation}
|\Psi_{v}\rangle^\mathrm{LCCSD} \equiv
\left( 1 + S_v^{(1)} + S_v^{(2)} +  C^{(1)} + C^{(2)}
\right) | 0_v \rangle \, .
\label{Eq:LCCSDwf}
\end{equation}

As discussed in the introduction, the cluster amplitudes can be found from solving
a proper analog of the eigen-value equation. We assume that these equations
are solved and in a typical application
we are faced with the necessity of computing matrix elements, Eq.~(\ref{Eq:melGeneral}),
between
two many-body wavefunctions $|\Psi_{v}\rangle$ and
$|\Psi_{w}\rangle$.
As demonstrated by \citet{BluJohLiu89}, so-called
disconnected diagrams~\cite{LinMor86} in the numerator and the denominator of
Eq.~(\ref{Eq:melGeneral}) cancel. Their final expression for the exact
matrix element reads
\begin{eqnarray}
\mathcal{M}_{wv}&=&\delta_{wv}\left(  Z^{\mathrm{core}}\right)_{\mathrm{conn}}+ \nonumber \\
& & \frac{\left(  Z_{wv}^\mathrm{val} \right)_{\mathrm{conn}}
}{\left\{  \left[  1+\left(  N_{v}^\mathrm{val} \right)_{\mathrm{conn}
}\right]  \left[  1+\left(  N_{w}^\mathrm{val} \right)_{\mathrm{conn}
}\right]  \right\}  ^{1/2}} \, ,
\label{Eq:Zconn}
\end{eqnarray}
where the matrix element $Z_{wv}$, Eq.(\ref{Eq:ZmelSeries}),
is split into core $Z^\mathrm{core}$ and
valence $Z_{wv}^\mathrm{val}$ contributions, the diagrams
comprising $Z^\mathrm{core}$ being independent of the
valence indexes.
The valence and core parts of the normalization
factor $N_{v}$ are defined in a similar fashion.
Notice that all the diagrams in Eq.~(\ref{Eq:Zconn})
must be rigorously connected as emphasized by  subscripts ``$\mathrm{conn}$''.
Since the total angular momentum of the closed-shell core is zero,
the core contribution $Z^{\mathrm{core}}$ vanishes for non-scalar (and pseudo-scalar)
operators and in the following discussion we will mainly focus on  $Z_{wv}^\mathrm{val}$.

\begin{figure}[h]
\begin{center}
\includegraphics*[scale=0.6]{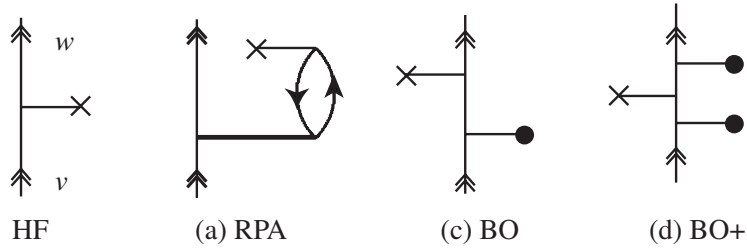}
\caption{ Dominant LCCSD contributions for the matrix elements. The double arrows
represent valence state, crosses represent matrix elements $z_{ij}$ and
heavy horizontal lines --- cluster amplitudes. In particular, the RPA diagram
involves valence doubles and BO diagram --- valence singles. Here and
below we do not draw the exchange variants for the diagrams.
 \label{Fig:ZSDRepresentative}}
 \end{center}
\end{figure}

\citet{BluJohLiu89} have employed the LCCSD parametrization for the
wavefunction~(\ref{Eq:LCCSDwf}) to derive  21 diagrams for $Z_{wv}^\mathrm{val}$
and 5 contributions to  $N_{v}^\mathrm{val}$. The LCCSD
contributions to $Z^{\mathrm{core}}$ can be found in Ref.~\cite{DerJohFri98}.
It is the goal of this paper to go beyond these linearized LCCSD contributions.
The LCCSD approximation will provide us with ``skeleton'' diagrams that will be ``dressed''
due to nonlinear CC terms.
We display  representative LCCSD diagrams in Fig.~\ref{Fig:ZSDRepresentative}.
In a typical calculation,
the dominant correction to the Hartree-Fock (HF) value
arise from RPA-type diagram (a) and Brueckner-orbital (BO) diagrams (c) and (d).
(We retain the original enumeration scheme of
Ref.~\cite{BluJohLiu89} for the diagrams.)  Here are the corresponding algebraic expressions
for these LCCSD contributions
\begin{eqnarray}
Z^{(\mathrm{HF})}_{wv} &=& z_{wv} \, , \nonumber \\
Z^{(\mathrm{a})}_{wv}  &=&  \sum_{ma} z_{am} \tilde{\rho}_{wmva} + \mathrm{h.c.s.} \, , \label{Eq:Zabare} \\
Z^{(\mathrm{c})}_{wv}  &=&  \sum_{m} z_{wm} \rho_{mv} + \mathrm{h.c.s.} \, , \nonumber \\
Z^{(\mathrm{d})}_{wv}  &=&  \sum_{mn} z_{mn} \rho^\ast_{mw} \rho_{nv} \, , \nonumber
\end{eqnarray}
where $\mathrm{h.c.s.}$ denotes hermitian conjugation of preceding term with a simultaneous
swap of the valence indexes, $w \leftrightarrow v$.

\section{Generating object $C^\dagger C$}
\label{Sec:CdagC}

At this point we have reviewed application
of the coupled cluster method to computing properties of univalent systems.
In the remainder of this paper we deal with mathematical object
\begin{equation}
\left( Z_{wv} \right)_{\mathrm{conn}} =
 \sum_{\lambda=0}^{\infty} \sum_{\mu=0}^{\infty} \frac{1}{\lambda!\mu!}
 \langle 0_c | a_w S^\dagger_w (C^\dagger)^\lambda \,
  Z \, (C)^\mu S_v a^\dagger_v | 0_c \rangle_{\mathrm{conn}} \, .
\label{Eq:ZmelSeriesUnivalent}
\end{equation}
As prescribed by the Wick theorem~\cite{LinMor86}, this expression may be simplified
by contracting creation and annihilation operators between various parts of this
expression. Very complex structures may arise, so
as a preliminary construct, consider a product $C^\dagger C$.
Using the Wick theorem, this product may be expanded into a sum of normal forms%
\begin{eqnarray}
\lefteqn{
C^{\dagger}C = \left(  C^{\dagger}C \right)_0  + \left(  C^{\dagger}C \right)_1 +
\left(  C^{\dagger}C \right)_2 + \cdots =} \\
& &c_0+\sum_{ij}c_{ij}\left\{  a_{i}^{\dagger}a_{j}\right\}  +\frac
{1}{2}\sum_{ij}c_{ijkl}\left\{  a_{i}^{\dagger}a_{j}^{\dagger}a_{l}%
a_{k}\right\}  +\cdots \nonumber
\label{Eq:CdagC}
\end{eqnarray}
Here notation $\left(  C^{\dagger}C \right)_k$ stands for $k$-body term.
The zero-body term $c_0$ does not have any free particle or hole lines and
would not contribute to connected diagrams of $Z_{wv}$.
One-body term  will lead to dressing of particle and hole lines, discussed
in Section~\ref{Sec:ParticleHole}. A part of two-body term will lead to
RPA-like dressing of LCCSD diagrams for matrix elements, as shown
in Section~\ref{Sec:ParticleHole}.

\section{Dressing particle and hole lines}
\label{Sec:ParticleHole}

\begin{figure}[h]
\begin{center}
\includegraphics*[scale=0.6]{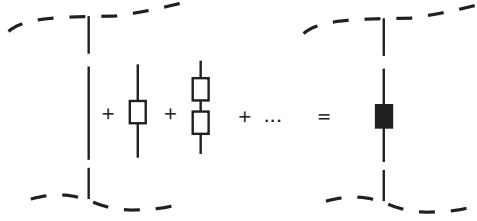}
\caption{ Schematic dressing of particle and hole lines in the CC diagrams
for matrix elements.
 \label{Fig:DressingGeneral}}
 \end{center}
\end{figure}

In this section we focus on one-body term of the product $C^\dagger C$,
Eq.~\eqref{Eq:CdagC}, and derive all-order insertions for particle and hole lines.
To this end it is useful to explicitly express the one-body term using particle and
core labels
\begin{eqnarray}
\left(  C^{\dagger}C \right)_1 &=&
-\sum_{ab} c_{ba}a_{a}a_{b}^{\dagger}+\sum_{mn}c_{mn}a_{m}^{\dagger}a_{n}+
\nonumber \\
&&
\sum_{ma} c_{ma}a_{m}^{\dagger}a_{a}+\sum_{ma}c_{am}a_{a}^{\dagger}a_{m} \, .
\label{Eq:OneBodyCCexpanded}
\end{eqnarray}
Topologically, the first term is an object where a free hole
line enters some (possibly very complex) structure from above and another hole line leaves below. The second
term has a similar structure but with particle lines.
In Fig.~\ref{Fig:DressingGeneral}, we draw these object as rectangles with
``stumps'' indicating where the particle or hole line is to be attached.
The remaining terms in Eq.~\eqref{Eq:OneBodyCCexpanded}
have both particle and hole lines involved; we will disregard these terms
in the following discussion.

First we  prove that given a certain CC diagram for matrix elements
we may ``dress'' all particle and hole lines as shown in Fig.~\ref{Fig:DressingGeneral}.
We start with a ``seed'' (``bare'') diagram coming from a certain set of contractions
in Eq.~\eqref{Eq:ZmelSeriesUnivalent}
\begin{equation}
\frac{1}{\lambda_0!\mu_0!}
 \langle 0_c | a_w S^\dagger_w (C^\dagger)^{\lambda_0} \,
  Z \, (C)^{\mu_0} S_v a^\dagger_v | 0_c \rangle_{\mathrm{seed}} \, .
\label{Eq:seed}
\end{equation}
As a next step consider a subset of terms of Eq.~\eqref{Eq:ZmelSeriesUnivalent},
constrained as $\lambda =  \lambda _{0}+n$, $\mu=  \mu _{0}+n$, $n=1,2,3,\ldots$.
In these terms carry out contractions within a product of $\lambda _{0}+n$ $C^\dagger$ operators
and $\mu _{0}+n$ $C$-operators
\[
\underset{\lambda _{0}+n}{\underbrace{C^{\dagger }\cdots C^{\dagger }}}
\underset{\mu _{0}+n}{~\underbrace{C \cdots C}} \, .
\]
Within this group there are $C^n_{\lambda_0+n} \times C^n_{\mu_0+n}$ ways
to pick out $n$ pairs of operators from the two sets, $C^n_k$ being binomial
coefficient. Once the two
strings of $n$ operators are chosen, there are $n!$ possible ways to
contract into pairs $\left(C^\dagger C\right)_1$ object. Finally, we
contract the $n$ resulting objects into a chain (see
Fig.~\ref{Fig:DressingGeneral}); there are $n!$ possible
combinations. Combining all these factors, we recover the original
factor $1/(\lambda_0! \mu_0!)$ in front of the seed
diagram~\eqref{Eq:seed}.

We may define a dressed particle-line insertion $\xi_{mn}$
in Fig.~\ref{Fig:DressingGeneral} as
\begin{eqnarray*}
\lefteqn{\sum_{mn} \xi_{mn} a^\dagger_m a_n = \sum_{mn} \delta_{mn} a^\dagger_m a_n +
 \left(C^\dagger C\right)_\mathrm{p-p} +} \\
&&\left( \left(C^\dagger C\right)_\mathrm{p-p} \left(C^\dagger C\right)_\mathrm{p-p} \right)_\mathrm{p-p}
+ \cdots \, ,
\end{eqnarray*}
where the subscript p-p denotes that we have to keep an insertion with a single incoming and
a single outgoing particle line, e.g.,
$ \left(C^\dagger C\right)_\mathrm{p-p}=\sum_{mn} c_{mn} a^\dagger_m a_n$.
Notice the absence of numerical factors in front of terms of the series; this fact follows from
the preceding discussion. Explicitly,
\begin{equation}
\xi_{mn} = \delta_{mn} + c_{mn} + \sum_r c_{mr} c_{rn} + \cdots \, .
\end{equation}
This series may be generated by iteratively solving an implicit equation
\begin{equation}
\xi_{mn} = \delta_{mn} + \sum_r c_{mr} \xi_{rn} \, .
\end{equation}
The very same argument holds for dressed insertions into the hole lines.

The derivation presented above can be generalized to include simultaneous dressing
of {\em all} particle-hole lines of a given diagram, including the inner lines
of the original ``bare'' object $\left(C^\dagger C\right)_1$ itself.
Below we illustrate our dressing scheme in case when the
cluster operator is truncated at single and double excitations.

\subsection{Singles-doubles approximation}

With the truncated cluster operator the hole-line insertion reads
\begin{equation}
c_{ba}    =\sum_{m}\rho_{mb}^* \rho_{ma}+
\frac{1}{2}\sum_{cmn}\tilde{\rho}^*_{mnbc}\tilde{\rho}_{mnac}
\end{equation}
and the particle-line insertion is
\begin{equation}
c_{mn}    =-\sum_{a}\rho^*_{na}\rho_{ma}-
\frac{1}{2}\sum_{abr}\tilde{\rho}^*_{nrab}\tilde{\rho}_{mrab} \, ,
\end{equation}
where we  introduced anti-symmetric quantities
$\tilde{\rho}_{mnab} = \rho_{mnab} - \rho_{nmab}$.

\begin{figure}[h]
\begin{center}
\includegraphics*[scale=0.4]{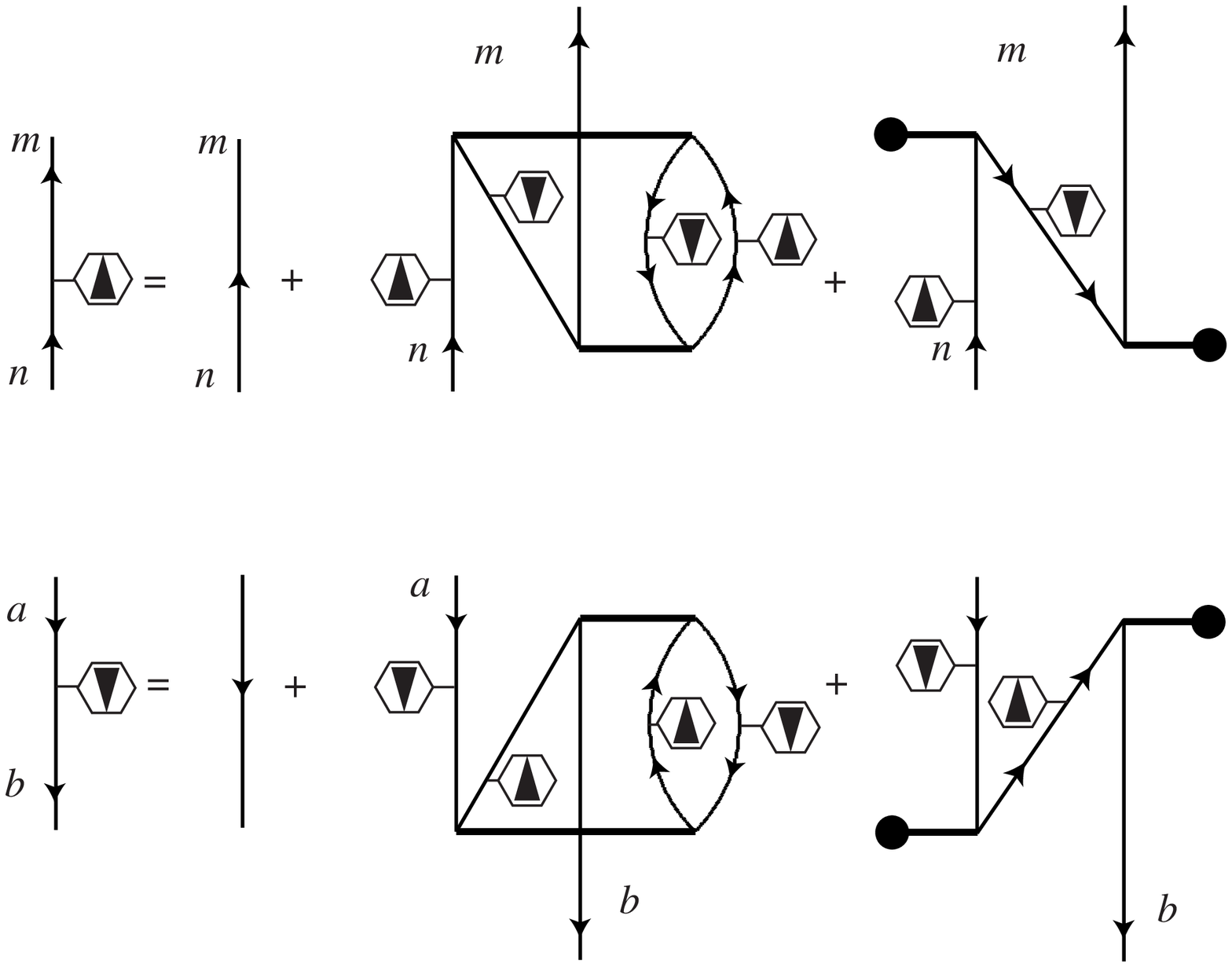}
\caption{ Dressing of particle and hole lines in the singles-doubles approximation.
The upper and lower panels  represent dressing of particle and hole lines respectively.
 \label{Fig:SDDressedInsertions}}
 \end{center}
\end{figure}

Diagrammatically,%
\[
c_{mn}a_{m}^{\dagger}a_{n}=%
\raisebox{-5ex}{\includegraphics[scale=0.4]{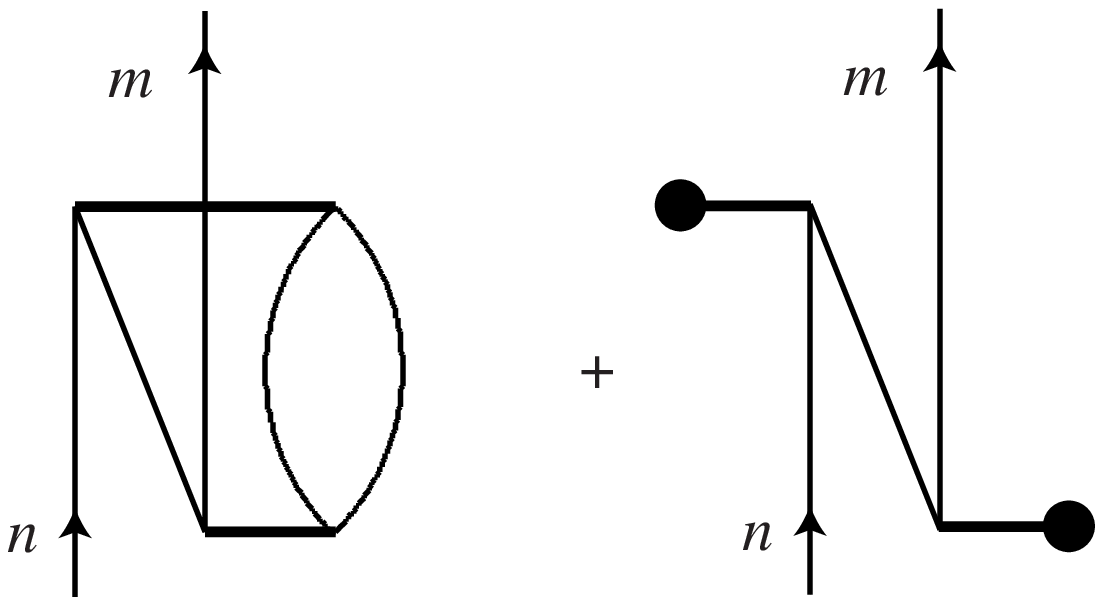}} \, .
\]
As discussed in the first part of this Section,
we may dress all the particle/hole lines according to the all-order
scheme in Fig.~\ref{Fig:SDDressedInsertions}.

Algebraically,
\begin{eqnarray}
\xi_{mn}  &  = &\delta_{mn}-\sum_{a}\rho_{ma}R_{na}^{\ast}-\sum_{abr}%
\tilde{\rho}_{mrab}R_{nrab}^{\ast} \, , \label{Eq:Xi}\\
\xi_{ba}  &  = &\delta_{ba}-\sum_{m}\rho_{mb}^{\ast}R_{ma}-\sum_{mnc}\tilde
{\rho}_{mnbc}^{\ast}R_{mnac} \, , \nonumber
\end{eqnarray}
where we introduced dressed core cluster amplitudes
\begin{eqnarray}
R_{na}^{\ast}  &  = & \sum_{b}\xi_{ba}\sum_{s}\rho_{sb}^{\ast}\xi_{sn} \, ,
\nonumber \\
R_{nrab}^{\ast}  & = & \sum_{t}\xi_{rt}\sum_{s}\xi_{ns}\sum_{c}\xi_{ca}
\sum_{d}\rho_{stcd}^{\ast}\xi_{db} \, .
\label{Eq:DressedCoreRho}
\end{eqnarray}
We solve Eq.~\eqref{Eq:Xi} iteratively
\begin{eqnarray}
\xi_{mn}^{\left(  i+1\right)  }  &=& \delta_{mn}-\sum_{a}\rho_{ma}%
R_{na}^{\left(  i\right)  \ast}-\sum_{abr}\tilde{\rho}_{mrab}R_{nrab}%
^{\left(  i\right)  \ast}\ ,\nonumber \\
\xi_{ba}^{\left(  i+1\right)  }  &=& \delta_{ba}-\sum_{m}\rho_{mb}^{\ast
}R_{ma}^{\left(  i\right)  }-\sum_{mnc}\tilde{\rho}_{mnbc}^{\ast}%
R_{mnac}^{\left(  i\right)  } ,
\label{Eq:InsertionsIterative}
\end{eqnarray}
where the dressed amplitudes $R_{\cdots}^{(  i)}$
are to be computed with coefficients $\xi_{\cdots}^{(i)}$ obtained at the previous step.

\begin{figure}[h]
\begin{center}
\includegraphics*[scale=0.6]{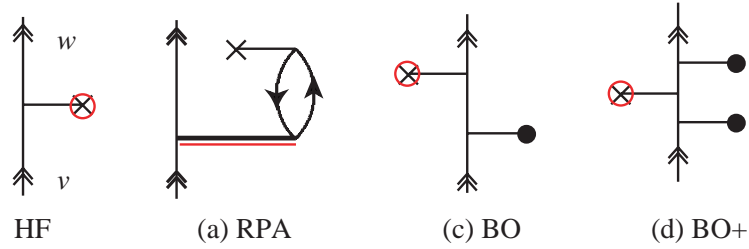}
\caption{ Dressing of particle-hole lines in the dominant
LCCSD diagrams, Fig.~\protect\ref{Fig:ZSDRepresentative}.
The circled crosses represent line-dressed matrix elements, Eq.~\protect\eqref{Eq:DressedMel},
and the double-lined valence cluster amplitude is the dressed amplitude,
Eq.~\protect\eqref{Eq:DressedValRho}.
 \label{Fig:ZSDRepresentativeDressed}}
 \end{center}
\end{figure}

With the calculated insertions $\xi$ we can ``upgrade'' the LCCSD diagrams
(compare Fig.~\ref{Fig:ZSDRepresentative} and Fig.~\ref{Fig:ZSDRepresentativeDressed}). To this end we introduce dressed matrix elements
$\bar{z}_{ij}$
and dressed valence cluster amplitudes $R_{mv}$ and $R_{mnva}$, similar to Eq.~\eqref{Eq:DressedCoreRho}
\begin{eqnarray}
\bar{z}_{mn} &= & \sum_{rs} \xi_{mr} z_{rs} \xi_{sn} \, , \nonumber \\
\bar{z}_{ab} & = & \sum_{cd} \xi_{ca} z_{cd} \xi_{db} \label{Eq:DressedMel} \, ,
\end{eqnarray}
\begin{eqnarray}
R_{mv}   &=& \sum_n \xi_{mn} \rho_{nv} \nonumber \, , \\
R_{mnva} &=& \sum_{brs} \xi_{nr} \xi_{ba} \xi_{sm}  \rho_{srvb} \, .
 \label{Eq:DressedValRho}
\end{eqnarray}
Notice that the  incoming valence line in the valence amplitudes $\rho_{mnav}$ and $\rho_{mv}$  is not dressed, since it does
not represent a free end. With these objects we may dress the LCCSD diagrams,
as shown in Fig.~\ref{Fig:ZSDRepresentativeDressed}.

Numerically, we rigorously computed the four dressed diagrams shown in
Fig.~\ref{Fig:ZSDRepresentativeDressed}.
In the remaining LCCSD diagrams, listed in Ref.~\cite{BluJohLiu89},
we have replaced the bare matrix elements with the dressed matrix elements,
Eq.~\protect\eqref{Eq:DressedMel}.
Notice that the
dressing of the Hartree-Fock diagram subsumes LCCSD diagrams
\begin{eqnarray}
Z^{(\mathrm{g})}_{wv} &=& -\sum_{ma} \rho_{ma}^* \rho_{wa} z_{mv} +\mathrm{h.c.s.} \, , \nonumber      \\
Z^{(\mathrm{t})}_{wv} &=& -\sum_{mnab} \rho_{mnba}^* \tilde{\rho}_{nwab} z_{mv} + \mathrm{h.c.s.}
\, . \label{Eq:LCCSDsubsumed}
\end{eqnarray}
from Ref.~\cite{BluJohLiu89}, so that these diagrams are to be
discarded in the present approach.
We postpone discussion of numerical results until Section~\ref{Sec:Numerics}.

\section{RPA-like dressing}
\label{Sec:RPA}
In this Section we continue with the systematic dressing of the coupled-cluster diagrams for
matrix elements based on the topological structure of the product $C^\dagger C$, Eq.~\eqref{Eq:CdagC}.
Here we focus on the two-body term of this object.
The insertion that generates the RPA-like chain of diagrams is
due to a two-particle/two-hole part of the object $\left( C^\dagger C \right)$,
\begin{equation}
 \left( C^\dagger C \right)_\mathrm{RPA} \equiv
 \sum_{mnab} \tilde{c}_{nbma}~ \left\{ a_{n}^{\dagger}a_{b}^{\dagger}a_{a}a_{m} \right\} \, ,
\label{Eq:RPAbare}
\end{equation}
where we used a symmetry property $c_{ijkl} = c_{jilk}$, and $\tilde{c}_{ijkl} = c_{ijkl} - c_{ijlk}$.
An analysis identical to that presented for line dressing in Section~\ref{Sec:ParticleHole}
leads us to introducing an RPA-dressed object $\mathcal{T}_{nbma}$,
\begin{eqnarray}
\lefteqn{
 \sum_{mnab}\tilde{\mathcal{T}}_{nbma}
 \left\{ a_{n}^{\dagger}a_{b}^{\dagger}a_{a}a_{m} \right\}
= -\sum_{mnab}\delta_{mn}\delta_{ab}
 \left\{ a_{n}^{\dagger}a_{b}^{\dagger}a_{a} a_{m} \right\} }  \label{Eq:RPASequence} \\
&+&
\left( C^\dagger C \right)_\mathrm{RPA} +
 \left(  \left(  C^{\dagger} C \right)_{\mathrm{RPA}}
\left(  C^{\dagger} C \right)_{\mathrm{RPA}}
 \right)_{\mathrm{RPA}}+ \cdots \nonumber \, ,
\end{eqnarray}
where subscript RPA specifies that the objects inside the brackets are to be contracted so that the result
has the same free particle/hole ends as the original bare object
$\left( C^\dagger C \right)_\mathrm{RPA} $, Eq.~\eqref{Eq:RPAbare}.
The sign of the leading term (with the Kronecker symbols) is chosen in such a way that
a product of this term with a particle-hole object like $z_{rc} a_c a^\dagger_r$ results in the
original object.

\begin{figure}[h]
\begin{center}
\includegraphics*[scale=0.7]{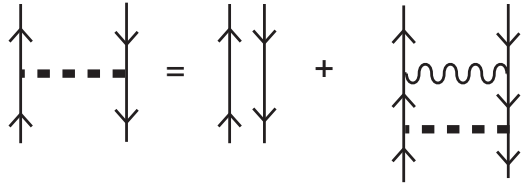}
\caption{ Graphical equation for RPA-dressed insertion. The dashed horizontal line
is the dressed insertion $\mathcal{T}_{nbma}$, while the bare object $c_{nbma}$
is represented by  a wavy line.
 \label{Fig:RPADressingEqn}}
 \end{center}
\end{figure}

A detailed consideration leads to an implicit equation for the RPA-dressed particle-hole
insertion
\begin{equation}
\tilde{\mathcal{T}}_{nbma} = - \delta_{mn}\delta_{ab} -
\sum_{rc} \tilde{c}_{ncra} \tilde{\mathcal{T}}_{rbmc} \, .
\label{Eq:RPAImplicit}
\end{equation}
This equation,
presented graphically in Fig.~\ref{Fig:RPADressingEqn}, can be solved iteratively.

\begin{figure}[h]
\begin{center}
\includegraphics*[scale=0.5]{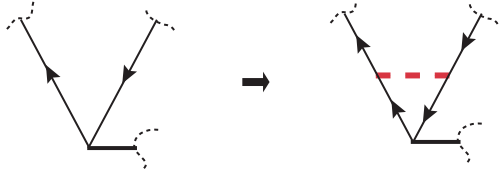}
\caption{ Dressing particle-hole vertex of a diagram with the RPA insertion,
Fig.~\protect\ref{Fig:RPADressingEqn}.
 \label{Fig:RPADressingVertex}}
 \end{center}
\end{figure}

The resulting insertion $\mathcal{T}_{nbma}$ may dress any particle-hole
vertex of a diagram as shown in Fig.~\ref{Fig:RPADressingVertex}.
As an example, consider dressing of particle-hole matrix elements of operator $Z$:
$z_{ma} a^\dagger_m a_a$. The equation for the dressed matrix element $ \bar{z}_{ma} $ may be derived
simply as
\begin{eqnarray*}
 \bar{z}_{ma}^\mathrm{RPA} & \equiv & - \sum_{nb} \tilde{\mathcal{T}}_{mbna} z_{nb} \, , \\
  \bar{z}_{ma}^\mathrm{RPA} &=& z_{ma} +
 \sum_{rc} \tilde{c}_{mcra}
 \sum_{nb} \tilde{\mathcal{T}}_{rbnc} z_{nb} = z_{ma} - \sum_{nb} \tilde{c}_{mbna} \bar{z}_{nb}^\mathrm{RPA} \, ,
\end{eqnarray*}
where we used  Eq.~\eqref{Eq:RPAImplicit} for the dressed object $\tilde{\mathcal{T}}_{mbna}$.
Finally we obtain a set of two equations
\begin{eqnarray*}
\bar{z}_{am}^\mathrm{RPA} &=&  z_{am} - \sum_{nb} \tilde{c}_{mbna}^* \bar{z}_{bn}^\mathrm{RPA} \, , \nonumber \\
\bar{z}_{ma}^\mathrm{RPA} &=&  z_{ma} - \sum_{nb} \tilde{c}_{mbna} \bar{z}_{nb}^\mathrm{RPA} \, .
\label{Eq:zRPAdressed}
\end{eqnarray*}
The resulting  equations resemble the traditional RPA formulae for dressed matrix elements
(see, e.g., Ref.~\cite{JohLiuSap96}), but do not couple $\bar{z}_{am}^\mathrm{RPA}$ and
$\bar{z}_{ma}^\mathrm{RPA}$. In addition,
the role of the residual Coulomb interaction of the traditional RPA formulae is played by the matrix elements of
$\left( C^\dagger C \right)_\mathrm{RPA}$ object (that is dominated by
second order in the Coulomb interaction, see Section~\ref{Sec:IV}).
In addition, we would like to emphasize that
the line-dressed matrix elements, Eq.~\eqref{Eq:DressedMel} also include matrix elements
between core and virtual orbitals; these are to be distinguished from the RPA-like matrix elements,
Eq.~\eqref{Eq:zRPAdressed}.

The above derivation is only valid when dressing a single isolated vertex.
More complex situation arises whenever a horizontal cross-cut through a diagram
produces several (not just two, as in Fig.~\ref{Fig:RPADressingVertex}) unquenched particle and hole lines.
Then the lower particle and hole lines of the object
$\left( C^\dagger C \right)_\mathrm{RPA}$ could be attached to the unquenched lines of the bare diagram in any order
and ``cross-dressing'' may occur. As an illustration, consider RPA-dressing of the valence double
contribution to the wavefunction $\sum_{mna} \rho_{mnva} a^\dagger_m a^\dagger_n a_a$. It arises,
for example, when the diagram $Z^{(a)}_{wv}$ (see Fig.~\ref{Fig:ZSDRepresentative}(a)) is cut across horizontally.
The valence double contains two equivalent vertexes $a^\dagger_m a_a$ and  $a^\dagger_n a_a$.
First we may attach the RPA object to the $m-a$
vertex and at the second step to the $n-a$ vertex. Apparently, this scenario is not covered by
Eq.~\eqref{Eq:RPASequence}, since it implies that all RPA insertions are attached to the same vertex.
Nevertheless, the  RPA-like dressing can be carried out in a straightforward fashion. To continue with
the illustration, the RPA-dressed valence double amplitude
may be defined as follows (compare to Eq.~\eqref{Eq:RPASequence} )
\begin{eqnarray}
\sum_{mna} \mathcal{R}_{mnva}^\mathrm{RPA} a^\dagger_m a^\dagger_n a_a &=&
\left(
\left[ 1 + \left( C^\dagger C \right)_\mathrm{RPA} +
 \left(  \left(  C^{\dagger} C \right)_{\mathrm{RPA}}
\left(  C^{\dagger} C \right)_{\mathrm{RPA}}
 \right)_{\mathrm{RPA}}+ \cdots \right] \times \right. \nonumber \\
&&\left.
\sum_{mna} \rho_{mnva} a^\dagger_m a^\dagger_n a_a
\right)_\mathrm{val. double} \, .
\label{Eq:GenRPADressValDouble}
\end{eqnarray}
Here subscript ``val. double'' indicates that we select a contribution having the
same free ends as the R.H.S. of the equation. The numerical factors in front of the
individual RPA contributions are derived similarly to the line-dressing factors
(see Section~\ref{Sec:ParticleHole}). Explicitly, we deal with a chain of diagrams
\begin{eqnarray}
\lefteqn{\mathcal{R}_{mnva}^\mathrm{RPA} = \rho_{mnva} -
\sum_{br}\tilde{c}_{nbra} \tilde{\rho}_{mrvb} +} \nonumber \\
& & \sum_{bcrs} \left(
 \tilde{c}_{ncsa} \tilde{c}_{sbrc}\tilde{\rho}_{mrvb}
-\tilde{c}_{ncsa} \tilde{c}_{mbrc} \tilde{\rho}_{srvb}
\right)+ \cdots \, .
\end{eqnarray}
Examining the structure of the above
expression we finally arrive at
\begin{equation}
\mathcal{R}_{mnva}^\mathrm{RPA}=\rho_{mnva}
-\sum_{sb} \tilde{c}_{nbsa}
\tilde{\mathcal{R}}_{msvb}^\mathrm{RPA} \, ,
\label{Eq:rhoRPAdressed}
\end{equation}
i.e., we have demonstrated how to dress the valence doubles.
This example can be easily generalized to several
equivalent vertexes: the Eq.~\eqref{Eq:GenRPADressValDouble} has to be rewritten using as the seed object
the underlying structure of the unquenched lines produced by a horizontal cross-cut through a diagram.

\subsection{Singles-doubles approximation}

Now we specialize our discussion of the RPA-like dressing
to the cluster operator truncated at single and double excitations.
In this CCSD approximation we obtain for the bare RPA-like insertion
\begin{equation}
\tilde{c}_{nbma} \approx - \rho_{mb}^* \rho_{na} -
\sum_{rc}\tilde{\rho}_{mrbc}^{\ast}\tilde{\rho}_{nrac} \, .
\label{Eq:CRPASD}
\end{equation}
Notice that one has to be careful when dealing with the first (singles $\times$ singles) term; it
is represented by a disconnected diagram and may produce undesirable disconnected diagrams for
matrix elements, Eq.~\eqref{Eq:Zconn}.

By substituting the CCSD insertion, Eq.\eqref{Eq:CRPASD},
into Eq.~\eqref{Eq:zRPAdressed} and Eq.~\eqref{Eq:rhoRPAdressed},
we immediately derive
expressions for the dressed matrix elements and the RPA-dressed valence doubles.
For example,
\begin{eqnarray}
\mathcal{R}_{mnva}^\mathrm{RPA}&=&\rho_{mnva}+
\sum_{sb}\left(
\sum_{rc}\tilde{\rho}_{nrac}\tilde{\rho}_{srbc}^{\ast}\right)
\tilde{\mathcal{R}}_{msvb}^\mathrm{RPA} \, , \label{Eq:RPAdressRho} \\
\bar{z}_{ma}^\mathrm{RPA} &=&  z_{ma} + \sum_{nb}
\left( \sum_{rc}\tilde{\rho}_{nrbc}^{\ast}\tilde{\rho}_{mrac} \right) \bar{z}_{nb}^\mathrm{RPA} \, .
\label{Eq:RPAdressMelSD}
\end{eqnarray}
Here we omitted a small contribution to  $\tilde{c}_{nbma}$, Eq.~\eqref{Eq:CRPASD},
from the product of core singles.
As discussed
in Section~\ref{Sec:IV} this contribution would arise in the higher orders (sixth order) of MBPT.

Practically, we notice that among the dominant LCCSD diagrams shown in
Fig.~\ref{Fig:ZSDRepresentative}, the particle-hole vertex occurs only in the RPA
diagram $Z_{wv}^{(a)}$. Focusing on this particular diagram,
the ``upgraded'' $Z_{wv}^{(a)}$ is simply
\begin{equation}
\left( Z^{(\mathrm{a})}_{wv} \right)_\mathrm{dress}  =
  \sum_{ma} z_{am} \tilde{\mathcal{R}}_{wmva}^\mathrm{RPA} + \mathrm{h.c.s.} \, .
  \label{Eq:ZaValRhoDressed}
\end{equation}
Notice that the use of the RPA-dressed matrix elements to upgrade the $Z_{wv}^{(a)}$ diagram,
such as
\begin{equation}
\left( Z^{(\mathrm{a})}_{wv} \right)_\mathrm{mel.\, dress}  =
  \sum_{ma} \bar{z}_{am}^\mathrm{RPA} \tilde{\rho}_{wmva} + \mathrm{h.c.s.} \,
\end{equation}
does not lead to the identical result, because it misses dressing of the $w-a$ vertex.
Moreover,  we found  numerically that dressing of both vertexes, as in more
general Eq.~\eqref{Eq:ZaValRhoDressed}, is equally important.

Finally, it is worth noting that with the RPA-like dressing scheme proposed here, the CCSD calculations
would recover the entire chain of RPA diagrams. However, if the calculations of the wavefunctions
are done using the linearized version of the coupled-cluster equations
(as in Refs.~\cite{BluJohLiu89,BluJohSap91,SafDerJoh98,SafJohDer99}),
a part of the RPA diagrams would still be missing~\cite{BluJohLiu89}. To summarize, the inclusion of the CCSD non-linear
terms in the CC equations is crucial for a fully consistent treatment of the RPA sequence.

\section{Comparison with the IV-order diagrams}
\label{Sec:IV}
Coupled cluster method can be straightforwardly connected with the direct
order-by-order many-body perturbation theory. In particular, for univalent systems,
when the calculations are carried out starting from the frozen-core Hartree-Fock
potential, the lowest-order contributions to the double excitation cluster amplitudes
are
\begin{eqnarray}
\rho_{mnva} &\approx&
 -\frac{g_{mnva}}{\varepsilon_m + \varepsilon_n-\varepsilon_a-\varepsilon_v} \, , \nonumber \\
\rho_{mnab} &\approx& -\frac{g_{mnab}}
{\varepsilon_m + \varepsilon_n-\varepsilon_a-\varepsilon_b} \, .
\label{Eq:RhoMBPTapprox}
\end{eqnarray}
Here $g_{ijkl}$ is the matrix element of the residual (beyond the HF potential)
Coulomb interaction between the electrons. It can be shown that while
the LCCSD method recovers all third-order diagrams for matrix elements, it starts
missing diagrams in the fourth order of MBPT. Our group has investigated
these 1,648 complementary IVth order diagrams in Refs.~\cite{DerEmm02,CanDer04}.
Among the diagrams  complementary to LCCSD matrix element contributions,
there are seven terms (class $Z_{1\times2}\left(  D_{nl}\right)$
in Ref.~\cite{DerEmm02} )  due to non-linear terms in the expansion of the CCSD
wavefunctions. Namely these $Z_{1\times2}\left(  D_{nl}\right)$ diagrams provide
the lowest-order approximation for the dressing scheme proposed here.

\begin{figure}[h]
\begin{center}
\includegraphics*[scale=0.5]{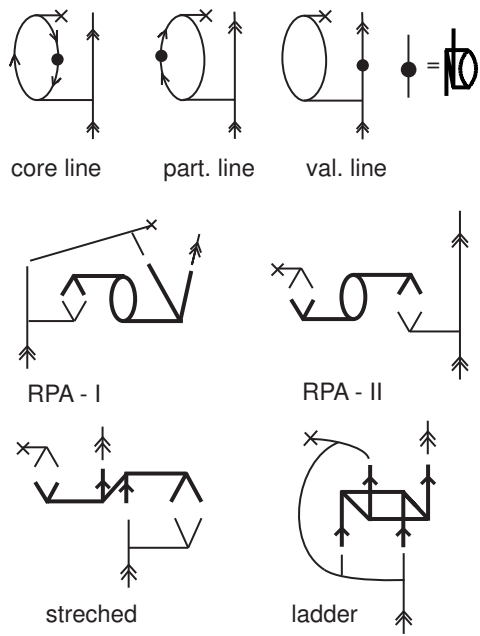}
\caption{Topological structure of fourth-order diagrams of class
$Z_{1\times2}\left(  D_{nl}\right)$ derived in
Ref.~\protect\cite{DerEmm02}. In these diagrams horizonal lines
denote Coulomb interaction and the topological objects coming from
the lowest-order approximation to $C^\dagger C$ are highlighted. The
upper three diagrams arise in the leading order of the line-dressing
scheme and two diagrams in the middle are related to the
RPA-dressing scheme.
\label{Fig:vsIVthOrder}}
\end{center}
\end{figure}

In Fig.~\ref{Fig:vsIVthOrder}
we explore the topological structure of  the $Z_{1\times2}\left(  D_{nl}\right)$ diagrams.
All these diagrams come from various ways of lowest-order dressing of $Z^{(\mathrm{a})}_{wv}$
diagram, Eq.~\eqref{Eq:Zabare}.
A comparison  shows that the present dressing approach recovers five
(three line-dressed and two RPA-dressed) out of seven IVth-order diagrams.
The missing diagrams are shown in the bottom row of Fig.~\ref{Fig:vsIVthOrder},
and we call them ``stretched'' and ``ladder'' diagrams, the names being derived from
the structure of the highlighted dressing insertions.
While the ladder diagram comes from the untreated two-body contribution to
$C^\dagger C$ object, Eq.~\eqref{Eq:CdagC},
the stretched diagram involves more complex {\em three}-body contribution to $C^\dagger C$.
Dressing with these two insertions can be carried out in a way similar to the
line- and RPA-like dressing schemes discussed in this paper and is beyond
the scope of the present analysis.

We verified that in the lowest-order MBPT approximation, Eq.~\eqref{Eq:RhoMBPTapprox}, our dressing formulae reproduce the
pertinent IVth-order expressions, explicitly presented in Ref.~\cite{DerEmm02}.
Furthermore, in Table~\ref{Tab:CsHFSIV} we list numerical values for individual contributions to
magnetic-dipole hyperfine structure constant (HFS) $A$ for ground state of $^{133}$Cs
and E1 transition amplitude for the principal $6p_{1/2}-6s_{1/2}$ transition in Cs.
Analyzing this Table we conclude that both line-- and RPA--like dressing are equally important.
Moreover, for these two particular matrix elements there is a partial cancelation between the
line-- and RPA-- dressed diagrams, so that were one of the dressings omitted,
the result would be misleading. We also observe that the untreated ``ladder'' diagram contributes
a negligibly small fraction of the total. At the same time the size of the
untreated ``stretched'' diagram indicates that (at least for the HFS constant) it is as important as the RPA-like
and line-dressed diagrams.

As to the numerics, the IVth-order calculations have been carried
out using relativistic B-spline  basis sets as described in
Ref.~\cite{JohBluSap88}.
We used a basis set of 25 out of 30 positive-energy ($\varepsilon_i > -m_e c^2$) pseudo-eigenfunctions
for each partial wave. Partial waves $s_{1/2}-g_{9/2}$ were included in the basis. The
summation over intermediate  core orbitals was limited to eight highest energy
core orbitals for the E1 amplitude and included all core orbitals for the HFS constant.
The reader is referred to paper~\cite{CanDer04} for a description of our IVth-order code.

\begin{center}%
\begin{table}[h]
\begin{tabular}{ldd}
\toprule
 Type &
\multicolumn{1}{c}{ $A(6s)$, MHz } &
\multicolumn{1}{c}{
$\langle 6s_{1/2}||D|| 6p_{1/2}\rangle, \mathrm{a.u.}$} \\
\colrule
 core line         & -2.0    &   7.0[-3]\\
 particle line     & -0.90   &   2.5[-3]\\
 valence line      & -0.92   &   0.2[-3]\\
 RPA-I             &  0.005  &  -9.3[-3]\\
 RPA-II            &  4.8    &  -0.05[-3]\\
 stretched         & -3.0    &   0.1[-3]\\
 ladder            &  0.05   &  -0.008[-3]\\
\hline
Total                & -1.9    &   0.41[-3]\\
\botrule
\end{tabular}
\caption{\label{Tab:CsHFSIV}
Breakdown of contributions to $Z_{1\times2}(D_{nl})$
class of diagrams for the hyperfine-structure (HFS)  constant $A$ for the $6s_{1/2}$ state
and $6s_{1/2}-6p_{1/2}$ electric-dipole transition amplitude for $^{133}$Cs
atom. The lowest-order DHF values are 1425.29 MHz for the HFS constant and
5.278 a.u. for the transition amplitude. Notation $x[y]$ stands for $x \times 10^{y}$.
}
\end{table}
\end{center}

We verified that numerical results for dressed all-order LCCSD diagrams (see Section~\ref{Sec:IV}) are consistent with the
values for the pertinent IVth-order diagrams.
As an example consider contributions to $A(6s)$ HFS constant.
Line-dressing modifies the LCCSD  $Z^{(\mathrm{a})}_{wv}$ diagram by $-4.3$ MHz
in a good agreement with a value of $-3.8$ MHz, the sum of the first three corresponding
IVth-order diagrams from Table~\ref{Tab:CsHFSIV}. Similarly, RPA-like dressing modifies
the $Z^{(\mathrm{a})}_{wv}$ diagram by $+4.4$ MHz, while the sum of IVth-order RPA diagram
from Table~\ref{Tab:CsHFSIV} is $4.8$ MHz.

\section{Numerical results and discussion}
\label{Sec:Numerics}
To reiterate discussion so far, we have developed the formalism of line- and RPA-like dressing
of the coupled-cluster diagrams for matrix elements. Further, we reduced our general formalism
to the case when the cluster operator is truncated at single and double excitation amplitudes.
We have also verified that in the lowest order we recover the relevant fourth-order diagrams
both analytically and numerically. In this Section we illustrate our  all-order dressing formalism
with numerical results.

We have carried
out relativistic calculations of the hyperfine-structure (HFS) constant $A$ for the $6s_{1/2}$ state
and $6s_{1/2}-6p_{1/2}$ electric-dipole transition amplitude for $^{133}$Cs
atom. It is worth noting that matrix elements of hyperfine interaction and electric dipole
operator allow one to access the quality of {\em ab initio}
wavefunctions both close to the nucleus and at intermediate values of
electronic coordinate. Such a test is essential for estimates of theoretical uncertainties
of calculations of parity-nonconserving (PNC) amplitudes. The {\em ab initio} PNC amplitudes
are key for high-accuracy probes of
new physics beyond the standard model of elementary particles with atomic
parity violation.

The results of calculations are presented in Tables~\ref{Tab:HFSsummary} and~\ref{Tab:Dsummary}.
In these tables we augment results of the previous all-order calculations~\cite{SafJohDer99} with
two types of new contributions: (i) all-order RPA- and line- dressing, outlined in
Sections~\ref{Sec:ParticleHole} and \ref{Sec:RPA}, and (ii) complementary
fourth-order contributions, so that the results are complete through the fourth-order
perturbation theory. Also, for the HFS constant, we incorporate the most recent values of
the Breit and radiative corrections~\cite{Der02,SapChe03}.

\begin{center}
\begin{table}[h]
\begin{tabular}{ld}
\toprule
\multicolumn{1}{c}{Contribution} &
\multicolumn{1}{c}{Value (MHz) } \\
\colrule
DHF             &     1425.29 \\
LCCSDpT, Coulomb&     2283.1 \\
Breit~\cite{Der02}    &        4.6 \\
VP+SE~\cite{SapChe03} &       -9.7  \\ 
\hline
LCCSDpT Reference  &      2278.0 \\[1ex]
%
\multicolumn{2}{c}{Dressing}  \\
$\Delta$(line-dress)   &  -11.0  \\
$\Delta$(RPA-dress)    &    4.4  \\
\hline
Dressing total         &  -6.7 \\
\multicolumn{2}{c}{Complementary IVth-order} \\
Triples\footnotemark[1]                   & +17.7 \\
$Z_{0 \times 3}(D_{nl})$  & -5.5 \\
$Z_{1\times2}\left(  D_\mathrm{nl}\right)$, stretched  & -3.0 \\
$Z_{1\times2}\left(  D_\mathrm{nl}\right)$, ladder     & +0.1 \\
\hline
$\Delta$(ZIV) total    &   +9.3 \\
\\
\hline
Final {\em ab initio}        &        2280.6 \\
Experiment   &        2298.2
\\
\botrule
\end{tabular}
\footnotetext[1]{ The VIth-order contributions from triple excitations are beyond
those treated in the LCCSDpT approximation.}
\caption{\label{Tab:HFSsummary}
Contributions to the  magnetic-dipole
hyperfine-structure constant $A$ of the ground $6s_{1/2}$ state of $^{133}$Cs.
}\end{table}
\end{center}

\begin{center}
\begin{table}[h]
\begin{tabular}{ld}
\toprule
\multicolumn{1}{c}{Contribution} &
\multicolumn{1}{c}{Value (a.u.) } \\
\colrule
DHF                 &   5.278   \\
LCCSD\protect\cite{SafJohDer99}
                    &   4.482   \\
LCCSDpT Reference   &   4.558    \\
\hline
\multicolumn{2}{c}{Dressing}  \\
$\Delta$(line-dress)   & 0.008 \\
$\Delta$(RPA-dress)    &-0.007  \\
\hline
Dressing corr. total   & -0.001 \\

\multicolumn{2}{c}{Complementary IVth-order} \\
Triples\footnotemark[1]                                & -0.043 \\
$Z_{0 \times 3}(D_{nl})$                               &  0.014 \\
$Z_{1\times2}\left(  D_\mathrm{nl}\right)$, stretched  &  1.0[-4] \\
$Z_{1\times2}\left(  D_\mathrm{nl}\right)$, ladder     & -8.6[-6] \\
\hline
$\Delta$(ZIV) total                                 &  -0.029   \\[1ex]

Final {\em ab initio}                   & 4.528                \\
\multicolumn{2}{c}{Experiment } \\
\citet{YouHilSib94}     & 4.5097(45)  \\
\citet{RafTanLiv99}     & 4.4890(65) \\
\citet{DerPor02}\footnotemark[2]        & 4.5064(47) \\
\citet{AmiDulGut02}\footnotemark[3]     & 4.5006(13) \\
\citet{AmiGou03}\footnotemark[4]        & 4.510(4) \\
\botrule
\end{tabular}
\footnotetext[1]{ The VIth-order contributions from triple excitations are beyond
those treated in the LCCSDpT approximation.}
\footnotetext[2]{ From van der Waals coefficient $C_6$ of the ground molecular state.}
\footnotetext[3]{ Photoassociation spectroscopy; this is the most accurate determination.}
\footnotetext[4]{ From static-dipole polarizability of $6s_{1/2}$ state with method of
Ref.~\cite{DerPor02}.}

\caption{\label{Tab:Dsummary}
Contributions to $\langle 6s_{1/2}||D|| 6p_{1/2}\rangle$
electric-dipole matrix element for Cs atom.
}
\end{table}
\end{center}

{\em LCCSDpT (perturbative triples) approximation.}
We depart from
the results of the coupled-cluster calculation described in Ref.~\cite{SafJohDer99}.
These are linearized coupled-cluster calculations, with the
wavefunctions truncated at single and double excitations from the reference
Slater determinant. In addition, following Ref.~\cite{BluJohSap91}, the perturbative
effect of triple excitations has been incorporated into the singles-doubles equation
(LCCSDpT method).
The main consequence of this perturbative treatment is that the resulting
valence removal energies are complete through the third order of perturbation
energies. There is also a substantial (a few per cent for Cs) improvement in the
accuracy of the resulting  LCCSDpT hyperfine constants over the LCCSD values.
At the same time the theory-experiment agreement for the E1 amplitudes significantly
degrades (see Table~\ref{Tab:Dsummary}):
while the LCCSD amplitudes differ by 0.4\% from 0.03\%-accurate experimental
data~\cite{AmiDulGut02}, the more sophisticated LCCSDpT matrix elements deviate
from measurements by as much as 1.3\%.
In other words, both LCCSD and LCCSDpT methods are poorly suited for  calculating
parity-non-conserving (PNC) amplitudes in $^{133}$Cs with uncertainty of a few 0.1\%.
It is one of the goals of this paper to establish a method that would provide a
consistent accuracy for both HFS constants and dipole matrix elements (and thus PNC amplitudes).

{\em Procedure.}
First we solved the relativistic LCCSDpT equations, as described in Ref.~\cite{SafJohDer99}.
With the computed cluster amplitudes we  calculated LCCSDpT  matrix elements and
recovered results published in Ref.~\cite{SafJohDer99}. Further, we
solved the line-dressing equations~\eqref{Eq:InsertionsIterative} and computed
the line-dressed cluster amplitudes and matrix elements.
The convergence rate was fast:
four iterations  were sufficient to stabilize the
norms of the line-dressed cluster amplitudes at a level of a few parts per million.
Finally, we solved the iterative equation for the RPA-dressed valence amplitudes,
Eq.\eqref{Eq:RPAdressRho}. It turned out to be very computationally intensive part of the scheme
and we iterated the equations only once.
We used the computed RPA-dressed valence amplitudes in
calculations of the dominant diagram $Z^{(a)}_{wv}$, see Eq.~\eqref{Eq:ZaValRhoDressed}.
In the remaining diagrams that involve particle-hole matrix elements, we employed
RPA-dressed matrix elements $\bar{z}^\mathrm{RPA}_{ij}$. A numerical iterative solution
of equations for $\bar{z}^\mathrm{RPA}_{ij}$, Eq.~\eqref{Eq:RPAdressMelSD}  required only a few iterations
to converge to seven significant figures.

{\em Line dressing.}
In Table~\ref{Tab:CsLineDress} we illustrate the importance of line dressing;
in this table we present differences between line-dressed and bare
LCCSDpT diagrams for the the hyperfine constant and E1 amplitude. The dressing of
the leading order HF diagram subsumes the LCCSD diagrams $Z^{(\mathrm{g})}_{wv}$
and $Z^{(\mathrm{t})}_{wv}$, Eq.~\eqref{Eq:LCCSDsubsumed}.
A direct calculation of these diagrams results in
$Z^{(\mathrm{g})}_{wv}+Z^{(\mathrm{t})}_{wv} = -22.9$ MHz for $A(6s)$ and $-1.75 \times 10^{-3}$ a.u. for
the E1 amplitude. These values are consistent with dressing-induced modifications
of the HF diagram ($-22.0$ MHz and $-1.7 \times 10^{-3}$  a.u., respectively) from Table~\ref{Tab:CsLineDress}. The modifications of the
$Z^{(\mathrm{a})}_{wv}$ diagram are consistent with the values of the pertinent
fourth-order diagrams, see Section~\ref{Sec:IV} and Table~\ref{Tab:CsHFSIV}.
A large dressing correction for HFS constant comes from the diagram $Z^{(\mathrm{c})}_{wv}$;
it is nominally a fifth-order
diagram. The relative importance of this diagram is not surprising since it is based upon
Brueckner orbitals (self-energy or core polarization effect).
As for the E1 amplitude, the line-dressing correction is dominated by $Z^{(\mathrm{a})}_{wv}$;
i.e., it is dominated by the fourth-order contribution. For the HFS constant, a
relative smallness of the line-dressing correction
to $Z^{(\mathrm{a})}_{wv}$ diagram  arises due to a delicate cancelation
of relatively large contributions from dressing of the core, particle and valence lines
of the diagram (see Table~\ref{Tab:CsHFSIV}).
Finally, in the bottom line of the Table~\ref{Tab:CsLineDress} we present a difference between
the line-dressed (all diagrams) and bare values. The dressing of the HF diagram plays a negligible
role here, since it is dominated by the diagrams already included in the LCCSDpT values.
The line dressing contributes at a sizable 0.5\% level to the HFS constant and at 0.2\% level to the
E1 amplitude.

\begin{center}%
\begin{table}[h]
\begin{tabular}{ldd}
\toprule
 Type &
\multicolumn{1}{c}{ $A(6s)$, MHz } &
\multicolumn{1}{c}{
$\langle 6s_{1/2}||D|| 6p_{1/2}\rangle, \mathrm{a.u.}$} \\
\colrule
$Z^{(\mathrm{HF})}_{wv}$ &  -22      &  -1.7[-3]  \\
$Z^{(\mathrm{a})}_{wv}$  &   0.04     &    9.3[-3]  \\
$Z^{(\mathrm{c})}_{wv}$  &  -8.6      &   -0.5[-3]  \\
$Z^{(\mathrm{d})}_{wv}$  &  -0.8      &   -3.6[-5]  \\
\hline
All diagrams              & -11      &   8.0[-3]   \\
\botrule
\end{tabular}
\caption{\label{Tab:CsLineDress}
Line-dressing induced modifications to dominant LCCSDpT
diagrams, Eq.~\eqref{Eq:Zabare}, for the HFS  constant $A$ for the $6s_{1/2}$ state
and $6s_{1/2}-6p_{1/2}$ E1 transition amplitude for $^{133}$Cs
atom.
}
\end{table}
\end{center}

{\em RPA-like dressing.}
Numerically dominant contribution due to the RPA-like dressing arises for $Z^{(\mathrm{a})}_{wv}$
diagram, where we used RPA-dressed valence amplitudes. The induced correction is as large as 0.2\% for both dipole amplitude and HFS constant.
The dressing of particle-hole matrix elements ($z_{ij} \rightarrow \bar{z}^\mathrm{RPA}_{ij}$) in
diagrams beyond $Z^{(\mathrm{a})}_{wv}$ played a relatively minor role, contributing at
a level of only 0.01\% for both test cases.

{\em Complementary fourth-order diagrams.} The LCCSDpT method misses certain many-body diagrams for
matrix elements starting from the fourth-order of MBPT. These complementary corrections in the
fourth order
come from triple and disconnected quadruple (or nonlinear double) excitations.
In Ref.~\cite{DerEmm02} these corrections were classified by the role of
triples and disconnected quadruples in the matrix elements
(i) an {\em indirect} effect of triples and disconnected quadruples on
single and double excitations lumped into class $Z_{0\times 3}$;
(ii) {\em direct} contribution  to matrix elements,  $Z_{1\times 2}$;
(iii) corrections to normalization, $Z_\mathrm{norm}$.
More refined classification reads
\begin{eqnarray}
\lefteqn{ \left( Z^{(4)}_{wv} \right)_\mathrm{non-LCCSD} =
Z_{1 \times 2}(T_v) + Z_{1 \times 2}(T_c) + } \nonumber \\
& &Z_{0 \times 3}(S_v[T_v]) + Z_{0 \times 3}(D_v[T_v]) +\\
& & Z_{0 \times 3}(S_c[T_c]) + Z_{0 \times 3}(D_v[T_c]) + \nonumber \\
& &Z_{1\times 2}\left( D_{\mathrm nl} \right) +
Z_{0\times 3}\left( D_{\mathrm nl} \right) + Z_\mathrm{norm}(T_v)  \nonumber \, .
\end{eqnarray}
Here we distinguished between valence ($T_v$) and core ($T_c$) triples and
introduced a similar notation for singles ($S$) and doubles ($D$).
Notation like $D_v[T_c]$ stands for effect of core triples ($T_c$) on
valence doubles $D_v$ through an equation for valence doubles.
The LCCSDpT method combines several diagrams
from $Z_{1 \times 2}(T_v)$ and $Z_{0 \times 3}(S_v[T_v])$ classes.
We removed these already included diagrams from the fourth-order
triples in Tables~\ref{Tab:HFSsummary}
and~\ref{Tab:Dsummary}.
Diagrams $D_\mathrm{nl}$ are
contributions of disconnected quadruples. As discussed in Section~\ref{Sec:IV}
namely one of such contributions,
$Z_{1\times 2}\left( D_{\mathrm nl} \right)$, provides the lowest-order
approximation to our all-order dressing scheme. In Tables~\ref{Tab:HFSsummary}
and~\ref{Tab:Dsummary} we added the contributions of untreated ``stretched'' and
``ladder'' diagrams of the $Z_{1\times 2}\left( D_{\mathrm nl} \right)$ class and
also from $Z_{0\times 3}\left( D_{\mathrm nl} \right)$ class. The latter contribution
would have been accounted for by solving the full (not linearized) CC equations.
In our large-scale fourth-order calculations  we have employed the code described in
Ref.~\cite{CanDer04}; all the formulas for a large number of diagrams
and the code have been generated automatically
using symbolic algebra tools. While the resulting fourth-order corrections from
triples are at the level of 1\%, we notice that there are certain noticeable cancelations
between various diagrams. Thus a complete all-order treatment of triples would be essential
for attaining the next level of theoretical accuracy.

{\em Hyperfine constant results.}
Details of calculation of the hyperfine constant $A(6s_{1/2})$ are presented in
Table~\ref{Tab:HFSsummary}. To clarify the role of correlations,
we first incorporate a number of small but important effects into the reference value:
Bohr-Weiskopf effect, Breit and radiative corrections.
The ``LCCSDpT, Coulomb'' value has been computed using the finite nuclear size,
both for determination of wavefunctions and computing matrix elements of hyperfine
interaction (this accounts for 0.5\%.) We also include
Breit corrections from Ref.~\cite{Der02}; these corrections
differ substantially from those incorporated in Ref.~\cite{SafJohDer99,BluJohSap91}
due to order-of-magnitude important correlation corrections.
Finally, radiative corrections to magnetic-dipole hyperfine-structure
constants for the ground state of alkali-metal atoms were computed
recently by \citet{SapChe03}. They found that the vacuum
polarization and self-energy (VP+SE) contribute as much as
0.4\% to the {\em ab initio} value. The reader should be careful with
adopting Breit values from Ref.\cite{SapChe03}, because these values
do not include correlation corrections (see Ref.~\cite{Der02} and references therein).
The final value, marked as ``LCCSDpT Reference'' deviates by 0.9\% from (exact) experimental
value.

Dressing corrections partially cancel, resulting in 0.3\% total dressing contribution. Fourth-order diagrams,
complementary to those already included in the LCCSDpT value are dominated by a contribution
due to triple excitations (0.8\%). We also include the ``stretched'' and ``ladder" IV-order diagrams
missed by our dressing scheme (see Section~\ref{Sec:IV}).
Almost all the  correlation corrections
are of similar sizes but of different signs, so the dressing and IV-th order corrections
cancel, so that the final correlation correction is only 0.1\%, just slightly improving
the theory-experiment agreement when compared with the ``LCCSDpT Reference'' value.
Our {\em ab initio} value for the HFS constant deviates by 0.8\% from the experimental value.

{\em Electric-dipole $6s_{1/2}-6p_{1/2}$ transition amplitude.} Details of
calculation for the dipole matrix element are compiled in Table~\ref{Tab:Dsummary}.
We do not include Breit and radiative corrections in that Table,
since the Breit interaction contributes only 0.02\% to this matrix element~\cite{Der02},
and radiative corrections are not known from the literature.

There were several high-accuracy experimental determinations of
$\langle 6p_{1/2} ||D|| 6s_{1/2} \rangle$ matrix element.
We list these matrix elements in the bottom of Table~\ref{Tab:Dsummary}.
In Refs.~\cite{YouHilSib94, RafTanLiv99} this matrix element
has been extracted from the measured
lifetime of the $6p_{1/2}$ state.  Determination of Ref.~\cite{AmiDulGut02}
is based on photoassociative spectroscopy of cold Cs atoms (i.e., inferred from  high-accuracy
measurement of molecular potentials). Another approach to extraction
of dipole matrix elements has been proposed by us in Ref.~\cite{DerPor02}:
we exploited an enhanced sensitivity of static electric-dipole polarizability $\alpha(0)$
of the ground state and van der Waals coefficient $C_6$ of the ground molecular
state to the matrix elements of principal transitions $\langle 6p_{3/2} ||D||
6s_{1/2} \rangle$ and $\langle 6p_{1/2} ||D|| 6s_{1/2} \rangle$.  Essential
to the extraction of individual matrix elements was
a high-accuracy ratio of these two dipole matrix elements measured
in Ref.~\cite{RafTan98}. Based on the proposed method~\cite{DerPor02},
the $\langle 6p_{1/2} ||D|| 6s_{1/2} \rangle$ matrix element has
been deduced from high-accuracy $C_6$ in Ref.~\cite{DerPor02}
and in Ref.~\cite{AmiGou03} it
has been inferred from $\alpha(0)$ measured in that work.
The most accurate matrix element comes from photoassociation
spectroscopy~\cite{AmiDulGut02}; their result has 0.03\% accuracy
and we will use that value below for calibrating  {\em ab initio} calculations.

The reference LCCSDpT E1 matrix element deviates from high-accuracy measurements by as much as 1.3\%.
The correlation corrections (dressing and fourth-order) computed by us improve the agreement
to about 0.6\% , i.e., the {\em ab initio} accuracy becomes comparable
to that for the HFS constant.
 An analysis of Table~\ref{Tab:Dsummary} shows that due to
cancelation of line- and RPA-like-dressing corrections the overall effect of dressing
is negligible for this transition amplitude. At the same time, the forth-order corrections due to
triple excitations beyond LCCSDpT triples are very large, almost 1\%. There corrections due
to residual fourth-order RPA corrections ($Z_{0\times3}(D_{nl})$) are also sizable,
and tend to decrease the effect of triples. Our forth-order calculation  demonstrates
that a full (beyond that of LCCSDpT) treatment of triple excitations improves the
accuracy of {\em ab initio} transition amplitudes.

\section{Conclusion}
\label{Sec:Conclusion}

The main two results of this work are: (i) development and application
of  all-order dressing formalism for matrix elements computed with coupled-cluster
method; (ii) first calculations of matrix elements for Cs complete
through the fourth order of many-body perturbation theory.

To reiterate, our dressing formalism is built upon a hierarchical expansion
of the product of clusters $C^\dagger C$ into a sum of $n$-body
insertions. We considered two types of insertions: particle/hole line
insertion
coming from the one-body part of the product and two-particle/two-hole
RPA-like insertion due to the two-body part. We  demonstrated
how to ``dress'' these insertions and formulated iterative equations.
Particular attention has been paid to the singles-doubles
truncation of the full cluster operator and we derived the
dressing equations for this popular approximation.
We have upgraded coupled-cluster diagrams for matrix elements
with the dressed insertions for univalent systems
and highlighted a relation to pertinent fourth-order diagrams. Finally, we illustrated
our formalism with relativistic calculations for Cs atom.

Our relativistic calculations also include a large number of fourth-order
diagrams complementary to LCCSDpT method (Linearized Coupled-Cluster Single-Doubles method
with perturbative treatment of Triples; it is the most sophisticated CC
approximation applied in relativistic calculations for Cs so far). The resulting
analysis is complete through the fourth-order of many-body perturbation theory.
We find that these complementary  diagrams substantially improve the theory-experiment
agreement for an important electric-dipole $6s_{1/2}-6p_{1/2}$ transition amplitude,
and slightly better the agreement for the hyperfine constant. We found sizable cancelations
between various fourth-order contributions; a full all-order treatment of triple and
disconnected quadruple excitations is desirable to further improve the theoretical
accuracy.

\acknowledgements

This work was supported in part by the National Science Foundation,
by the NIST precision measurement grant, and by the Russian
Foundation for Basic Research under Grant No.\ 04-02-16345-a.


\end{document}